\documentclass[11pt]{article}
\usepackage{amsmath}
\usepackage{amsfonts}
\usepackage{amssymb}
\usepackage{fontenc}

\newtheorem{theo}{Theorem}[section] 
\newtheorem{prop}[theo]{Proposition}

\def\be{\begin{equation}}
\def\ee{\end{equation}}
\def\bea{\begin{eqnarray}}
\def\eea{\end{eqnarray}}

\def\scri{{\cal{I}}}
\begin{document}
\title{The St\"utzfunktion and the Cut Function\footnote{This is a corrected, revised and updated 
version of a paper which originally appeared in {\it{Recent Advances in General Relativity}} eds. A I Janis and J R Porter, Einstein Studies vol. 4, Birkha\"user 1992}}
\author{Paul Tod}
\date{}
\maketitle
\begin{abstract}
I review some standard theory of convex bodies in $\mathbb{R}^3$ and rephrase it in a formalism of Ted Newman to show the relation between the St\"utzfunktion of the 
former theory and the cut function introduced by Ted. This leads to a conjectured inequality for space-like two-spheres in Minkowski space that generalises Minkowski's inequality 
and is implied by Penrose's cosmic censorship hypothesis.
\end{abstract}
\section{Introduction}

The work described in this paper arises from a problem posed to me by Ted Newman during my first visit to the University of Pittsburgh in the mid-1970s. It turns out that this problem can be solved 
by Newman-style methods and that it leads on to making interesting connections with other areas of Ted's work.

The problem is as follows: given a cut $\Sigma$ of the future-null infinity $\scri^+$ of Minkowski space $\mathbb{M}$, how do you reconstruct a space-like 2-surface $S$ inside 
$\mathbb{M}$ such that $\Sigma$ is the intersection with $\scri^+$ of $\dot{J}(S)$, the boundary of the future of $S$? This is related to a version of the ``fuzzy point'' idea which 
was current at that time: if $\Sigma$ is a cut of the $\scri^+$ of a non-flat but asymptotically-flat space-time ${\mathcal{M}}$ arising from a point $p$ in ${\mathcal{M}}$ (known then as a {\it{light-cone cut}}) then, when 
transferred to the $\scri^+$ of $\mathbb{M}$, $\Sigma$ will determine a null hypersurface ${\mathcal{N}}$ which does not converge to a point; however ${\mathcal{N}}$ may nearly converge to a point 
and may determine small 2-surfaces $S$ which are nearly points, or are fuzzy points. If so, then by taking all possible $\Sigma$ for all possible $p$ in ${\mathcal{M}}$, one might obtain a representation of 
the curved space-time ${\mathcal{M}}$ as fuzzy points in the flat space-time $\mathbb{M}$.

\medskip

The plan of this paper is as follows:

In section 2 I discuss convex bodies in $\mathbb{R}^3$. The theory of convex bodies centres on the {\it{St\"utzfunktion}} or support-function, which I'll anglicise as stutzfunction, and I review some of this theory.

In section 3, I turn to Minkowski space and identify the relation between the stutzfunction of a convex body and the cut-function which the boundary of the future of the body defines at $\scri^+$. This effectively 
solves the problem posed above, and it also illuminates some of the theory in section 2.

In section 4, I sketch some further developments of the theory of convexity for 2-surfaces in Minkowski space. These include an approach to an inequality found by Gibbons and Penrose \cite{rp1}, \cite{g1}, as a prediction of the cosmic censorship hypothesis.

\medskip

It gives me great pleasure to dedicate this paper to Ted Newman in his 60th year\footnote{Note the previous footnote.}, to acknowledge his long-standing and beneficial influence and to record my debt and gratitude to him.

\section{Convex bodies in $\mathbb{R}^3$}
In this section, I develop some standard theory of convex bodies in $\mathbb{R}^3$ following \cite{bla} and \cite{egg}, but with the kind of formalism that I learned from Ted Newman.

We may define a {\it{convex body}} $B$ in $\mathbb{R}^3$ to be a closed body such that, if $p,q$ are two points of $B$ then the line segment $tp+(1-t)q$ for $0\leq t\leq 1$ lies entirely in $B$. Then a {\it{convex surface}} $S$
is the surface of a convex body.

We define the Gauss map in the familiar way: choose an orthonormal triad and parametrise a unit vector $\ell$ by spherical polars as
\be\label{1}
\ell=\ell(\theta,\phi)=(\sin\theta\cos\phi,\sin\theta\sin\phi,\cos\theta),\ee
in the triad. Thus $\ell$ corresponds to a point on the unit sphere $S^2$. Given a choice of $\ell$, that is a choice of $(\theta,\phi)$, take the plane with normal $\ell$ that is tangent to the convex surface $S$, with $\ell$ 
the outward normal. If this happens at 
$p\in S$ then the Gauss map from $S$ to $S^2$ takes $p$ to the point labelled $(\theta,\phi)$ on the unit sphere. For a smooth, strictly convex body, the Gauss map is one-one as we shall see. In that case we have introduced coordinates $(\theta,\phi)$ 
on $S$. The tangent plane to $S$ at $p$ has the equation
\be\label{2}
{\bf{x\cdot\ell}(\theta,\phi)}=H(\theta,\phi),\ee
where $H(\theta,\phi)$ is the perpendicular distance from the origin (which we'll assume to be inside $S$) to the tangent plane. A knowledge of $H(\theta,\phi)$ 
determines $S$ as an envelope of tangent planes and $H$ is the {\it{St\"utzfunktion}} 
\cite{bla} or support function, which we'll call the stutzfunction.

To obtain a parametric expression for the surface $S$ we can solve (\ref{2}) and its derivatives for $(x,y,z)$. To this end, we introduce the Newman-Penrose operator `eth' \cite{np2} defined on a spin-weight $s$ function $\eta$ by
\[
\eth\eta:=\frac{1}{\sqrt{2}}(\sin\theta)^s\left(\frac{\partial}{\partial\theta}+\frac{i}{\sin\theta}\frac{\partial}{\partial\phi}\right)(\sin\theta)^{-s}\eta,
\]
and define
\[{\bf m}=\eth\ell=\frac{1}{\sqrt{2}}(\cos\theta\cos\phi-i\sin\phi,\cos\theta\sin\phi+i\cos\phi,-\sin\theta)\]
\[\overline{\bf m}=\overline{\eth}\ell=\frac{1}{\sqrt{2}}(\cos\theta\cos\phi+i\sin\phi,\cos\theta\sin\phi-i\cos\phi,-\sin\theta).\]
These have $s=1,-1$ respectively and, by differentiating again,\
\[
\eth {\bf m}=0=\overline{\eth}\overline{\bf m},\;\;\;\overline{\eth} {\bf m}=\eth\overline{\bf m}=-\ell.\]
The positive-definite metric of $\mathbb{R}^3$ can be written
\be\label{5}
\delta=\ell\ell+{\bf m\overline{m}}+{\bf \overline{m}m}.
\ee
Note also, as usual, that
\be\label{6}
(\eth\overline{\eth}-\overline{\eth}\eth)\eta=-s\eta,
\ee
when $\eta$ has spin-weight $s$.

To obtain the convex surface parametrically we must solve (\ref{2}) simultaneously with 
\[{\bf{x}}\cdot {\bf m}=\eth H,\;\;\;{\bf{x}}\cdot \overline{\bf m}=\overline{\eth} H,\]
which with the aid of (\ref{5}) can be solved to give 
\be\label{7}
{\bf x}={\bf x}(\theta,\phi)=H\ell+\overline{\eth}H{\bf m}+\eth H\overline{\bf m}.
\ee
This gives an explicit parametrisation of $S$. Using the standard theory of surfaces in $\mathbb{R}^3$ (see e.g. \cite{doc}) we find the area element of $S$ to be
\be\label{9}
dA=\left((H+\eth\overline{\eth}H)^2-\eth^2H\overline{\eth}^2H\right)\sin\theta d\theta d\phi.\ee
By general theory, the Jacobian of the Gauss map is the Gauss curvature $k$ so that
\be\label{10}
k^{-1}=R_1R_2=(H+\eth\overline{\eth}H)^2-\eth^2H\overline{\eth}^2H
\ee
in terms of principal radii of curvature $R_1,R_2$. A similar calculation gives the mean curvature $h$:
\[h=\frac12(R_1^{-1}+R_2^{-1})=(H+\eth\overline{\eth}H)k,\]
so that
\[hdA=(H+\eth\overline{\eth}H)\sin\theta d\theta d\phi.\]
For use below note that then
\be\label{h}
\int_ShdA=\int_{S^2}H\sin\theta d\theta d\phi,\ee
since $\eth\overline{\eth}$ is a constant multiple of the 2-sphere Laplacian, so the second term integrates to zero.

For strict convexity we require $R_1,R_2>0$ which is equivalent to $h,k>0$. Since necessarily $h^2\geq k$ it is sufficient to require $k>0$ ($h$ will be positive somewhere on $S$ since we are using the outward normal). 
Finally, since this is the Jacobian of the Gauss map, the Gauss map is one-one and onto precisely for smooth, strictly convex surfaces. This imposes restrictions on $H$ which we shall discuss. First we note how 
simple the Gauss-Bonnet theorem is in this context:
\begin{prop}{\bf The Gauss-Bonnet Theorem}

For a strictly convex surface $S$
\[\int_S kdA=4\pi.\]
\end{prop}

{\bf Proof}

From (\ref{9}) and (\ref{10}), $kdA=\sin\theta d\theta d\phi$.

\medskip

\noindent Now what conditions do we require $H$ to satisfy for it to be the stutzfunction of a smooth, strictly convex surface? We need $H$ positive and in (\ref{10}) we want the right-hand-side to be positive. If it fails to be positive then the surface enveloped by (\ref{2}) will have cusps. However if $k$ from (\ref{10}) 
is not positive then it can be made positive by adding a positive constant to $H$.

The process of adding a positive constant to $H$ is an interesting transformation that changes a convex surface $S$ into another, $S'$, which is {\it parallel} to it in the sense of \cite{ste}. In that 
reference, the idea is motivated by imagining rolling a sphere $K$ of constant radius $s$ over 
the surface $S$. The locus of the centre of $K$ defines $S'$. Equivalently one moves the centre of $K$ over $S$ and takes $S'$ to be envelope swept out by $K$. In this second form, one sees a connection with the 
idea of {\it Huyghens' secondary wavelets} which will reappear in section 3.

Given a convex surface $S$, one can consider a sequence of surfaces parallel to $S$ with larger and larger separations $s$. In this way one arrives at the following string of theorems.
\begin{prop}{\bf Steiner's Theorem}

Along such a sequence, the area is given by
\be\label{11}
A(s)=A+2sM+4\pi s^2,
\ee
while the volume contained is
\be\label{12}
V(s)=V+sA+s^2M+\frac{4}{3}\pi s^3.\ee
Here $A$ and $V$ are the area and volume of $S$ and $M$ is the integral of mean curvature:
\be\label{13}
M=\int_ShdA.\ee
\end{prop}

{\bf Proof}

Clearly the surface parallel to $S$ at distance $s$ has stutzfunction $H+s$. Substitute into (\ref{9}) and expand in powers of $s$ to obtain (\ref{11}) (using (\ref{h}) along the way). Integrate (\ref{11}) to obtain (\ref{12}).

\medskip

\noindent Along a sequence of parallel surfaces, the surfaces should become ``rounder''. This intuitive feeling is made precise in the following (which I won't prove):
\begin{prop}{\bf The Brunn-Minkowski Theorem}

 Define $R(s)=(V(s))^{1/3}$ then $R$ is convex in $s$, in that $\frac{d^2R}{ds^2}\leq 0$.
\end{prop}
\noindent Take this to be true and calculate the derivative, then for positive $s$:
\be\label{14}
(6MV-2A^2)+2s(12\pi V-AM)+2s^2(4\pi A-M^2)\leq 0.\ee
From this we may deduce {\it Minkowski's inequality}
\be\label{15}
M^2\geq 4\pi A,\ee
as well as the {\it {isoperimetric inequality}}
\be\label{16}
36\pi V^2\leq A^3.\ee
Although (\ref{15}) follows from (\ref{14}) there is a straightforward direct proof due to Blaschke and using the stutzfunction (see \cite{bla}):

\medskip

\noindent{\bf Blaschke's proof of Minkowski's Inequality}

\medskip

From (\ref{9}),(\ref{6}) and integration by parts
\[A=\int(H^2-\eth H\bar{\eth}H)\sin\theta d\theta d\phi\]
while from (\ref{h})
\[M=\int H\sin\theta d\theta d\phi.\]
 Set $H=H_0+H_1$ where $\int H_1\sin\theta d\theta d\phi=0$ and $H_0$ is constant then
 \[M=4\pi H_0,\;\;A=4\pi H_0^2+\int(H_1^2-\eth H_1\bar{\eth}H_1)\sin\theta d\theta d\phi.\]
 It follows by expanding $H_1$ in spherical harmonics that the integral contribution to $A$ is strictly negative unless $H_1$ is a combination of $\ell=1$ spherical harmonics. This case corresponds to a sphere with a translated 
 origin, so that (\ref{15}) is proved, with equality only for a round sphere.
 
 \medskip
 
 To conclude this section, I shall record another way of obtaining the surface $S$ from the stutzfunction $H$. Define
 \[\hat{H}(r,\theta,\phi)=rH(\theta,\phi)=F(\hat{x},\hat{y},\hat{z}),\]
 where $\hat{x},\hat{y},\hat{z}$ are expressed in terms of spherical polar coordinates $r,\theta,\phi$ in the usual way. Then (\ref{7}) is equivalent to the parametrisation given by
 \[x=\frac{\partial F}{\partial \hat{x}},\;\;y=\frac{\partial F}{\partial \hat{y}},\;\;z=\frac{\partial F}{\partial \hat{z}},\]
 where, after the differentiation, $\hat{x},\hat{y},\hat{z}$ are again eliminated.
 \section{Stutzfunction and cut function}
 In Minkowski space $\mathbb{M}$ we introduce the null tetrad
 \[
 \ell^a=(1,\ell),\;\;m^a=\eth\ell^a=(0,{\bf m}),\;\;\overline{m}^a=\overline{\eth}\ell^a=(0,\overline{\bf m}),\;\;n^a=\frac12(1,-\ell),\]
 with $\ell,{\bf m},\eth$ as in section 2.
The Minkowski metric can then be written
\[\eta^{ab}=2\ell^{(a}n^{b)}-2m^{(a}\overline{m}^{b)}.\]
 Note that
 \[
 \overline{\eth} m^a=(0,-\ell)=-\frac12\ell^a+n^a.\]
 Define the unit time-like vector
 \be\label{24}
 t^a=(1,{\bf 0})\ee
 so that also $\eta_{ab}t^a\ell^b=1$, and introduce advanced null polar coordinates $(u,r,\theta,\phi)$ by
 \[
 x^a=ut^a+r\ell^a(\theta,\phi)\]
 (see e.g. \cite{knt}). Then $(u,\theta,\phi)$ are coordinates on $\scri^+$ which is located at $r=\infty$.
 
 A cut of $\scri^+$ is defined by a function
 \[u=V(\theta,\phi)\]
 where $V$ can conveniently be called \emph{the cut function} for the cut. If we choose an arbitrary point $p$ with coordinates $x_0^a$ and a null-vector $\ell^a(\theta_0,\phi_0)$ at $p$ then 
 the null geodesic from $p$ in the direction of $\ell^a(\theta_0,\phi_0)$ meets $\scri^+$ at 
 \be\label{27}u=x_0^a\ell_a(\theta_0,\phi_0),\;\;\;\theta=\theta_0,\;\;\;\phi=\phi_0.\ee
Now suppose we are given a convex surface $S$ in the form
\[x^a=(0,{\bf x}(\theta,\phi))\]
with ${\bf x}(\theta,\phi)$ determined by a stutzfunction $H$ according to (\ref{7}). The boundary of the future of $S$, $\dot{J}(S)$, is ruled by null geodesics that meet $S$ orthogonally. From the definitions so 
far made, the outward null normal at the point $p$ of $S$ labelled by $(\theta_0,\phi_0)$ is $\ell^a(\theta_0,\phi_0)$ and the null geodesic from $p$ in this direction meets $\scri^+$ at $u$ given by 
(\ref{27}). As $p$ runs over $S$, the cut $\Sigma=\dot{J}(S)\cap\scri^+$ is generated with the cut function
\be\label{29} u=x^a\ell_a(\theta,\phi)=-{\bf x\cdot}\ell(\theta,\phi).\ee
Comparing (\ref{29}) with (\ref{2}) we conclude that \emph{the cut function is minus the stutzfunction}.

\medskip

Conversely, if we are given the cut $\Sigma$ and its cut function $V$ then (\ref{29}) determines a null hypersurface $\mathcal{N}$
\[x^a\ell_a(\theta,\phi)=V(\theta,\phi),\]
which, with its angular derivatives, we can solve for $\mathcal{N}$ parametrically as
\[x^a=x^a(\lambda,\theta,\phi)=-Hn^a+\overline{\eth} Hm^a+\eth H\overline{m}^a+\lambda \ell^a\]
for arbitrary real $\lambda$. Now intersecting $\mathcal{N}$ with hypersurfaces of constant $t$, i.e. hypersurfaces orthogonal to $t^a$ in (\ref{24}), gives a sequence of parallel surfaces in the sense of section 2. 
This shows how the parallel-surface idea is related to Huyghen's secondary wavelets: if a convex surface is momentarily lit up then the resulting expanding (out-going) wave-front traces out a sequence of surfaces parallel to the first.

Of course this converse is incomplete in the following sense: if what we are given is just the cut function then we can define the null surface $\mathcal{N}$ but we cannot fix a unique value of $\lambda$ to represent 
the 2-surface $S$ without some extra input. If we are trying to make precise the fuzzy-point idea then we might want to pick out an instant of {\it minimum volume} or of {\it best focus}. This could also involve boosting the cut function or 
considering a different set of constant-time hypersurfaces.
\section{Further developments}
In this last section I describe some attempts to carry over other parts of the theory of convex 2-surfaces into Minkowski space. I shall work with the GHP formalism \cite{ghp} and omit proofs.

A space-like 2-surface $S$ in ${\mathbb{M}}$ defines a pair of future-pointing null normals $\ell^a,n^a$ (where we shall take $\ell^a$ to be the outward normal and $n^a$ the inward normal) or equivalently a normalised spinor 
dyad $(o^A,\iota^A)$. The second fundamental form of $S$ is coded by the GHP formalism into weighted scalars $(\rho,\rho',\sigma,\sigma')$ (see e.g. \cite{t1} for an account of this). In terms of these, I shall 
say that $S$ is
\[\mbox{future convex iff }\rho<0,\;\;\;\rho^2-\sigma\overline{\sigma}>0\]
\[\mbox{past convex iff }\rho'>0,\;\;\;\rho'^2-\sigma'\overline{\sigma}'>0.\]
We recall that the Gauss curvature of $S$ is twice the real part of the \emph{complex curvature} $Q=-\rho\rho'+\sigma\sigma'$ \cite{pr}.
\begin{prop}

 If $S$ is future and past convex then the Gauss curvature of $S$ is everywhere positive.
 
\end{prop}

{\bf Proof:} This is elementary since
\[k=-2\rho\rho'+\sigma\sigma'+\overline{\sigma}\overline{\sigma}'\geq-2\rho\rho'-2|\sigma\sigma'|\]
and
\[(-\rho\rho'-|\sigma\sigma'|)^2=(\rho^2-\sigma\overline{\sigma})(\rho'^2-\sigma'\overline{\sigma}')+(\rho|\sigma'|+\rho'|\sigma|)^2.\]

\noindent The quantities occuring in the above definitions of convexity arise in the various Gauss maps that can be defined for $S$. If we choose and fix a constant normalised spinor dyad $(\alpha^A,\beta^A)$ then we 
can define a \emph{future Gauss map} by
\[f:S\rightarrow {\mathbb{CP}}^1;\;\;\;\;p\mapsto \zeta=\frac{o_A\alpha^A}{o_A\beta^A}\]
and a \emph{past Gauss map} by
\[f:S\rightarrow {\mathbb{CP}}^1;\;\;\;\;p\mapsto \eta=\frac{\iota_A\alpha^A}{\iota_A\beta^A}.\]
Equivalently, these express $o^A$ and $\iota^A$ in terms of $(\alpha^A,\beta^A)$  as
\[o^A=\lambda\frac{(\alpha^A+\zeta\beta^A)}{(1+\zeta\overline{\zeta})^{1/2}},\;\;\;\iota^A=\mu\frac{(\alpha^A+\eta\beta^A)}{(1+\eta\overline{\eta})^{1/2}},\]
where $\lambda,\mu$ are not fixed by the Gauss maps, but note that
\[t^a\ell_a=\frac{1}{\sqrt{2}}\lambda\overline{\lambda},\;\;\;t^an_a=\frac{1}{\sqrt{2}}\mu\overline{\mu},\]
where $t^{AA'}= \frac{1}{\sqrt{2}}(\alpha^A\overline{\alpha}^{A'}+\beta^A\overline{\beta}^{A'})$, so that $t^a$ is a unit time-like vector determined by the chosen tetrad. (These Gauss maps are similar to but 
different from those defined by Kossowski \cite{kos}.)

There is a third Gauss map, conveniently called the \emph{complex Gauss map} which can be defined by
\[2o^{(A}\iota^{B)}=\zeta\alpha^A\alpha^B+2\eta \alpha^{(A}\beta^{B)}+\xi\beta^A\beta^B\]
where this $\zeta,\eta$ are to be distinguished from the previous. This maps $S$ to the complex quadric $\mathbb{Q}$ defined by
\[\zeta\xi-\eta^2=1\]
in $\mathbb{C}^3$ (this Gauss map has also been considered by Roger Penrose).

The images of the future and past Gauss maps carry volume forms $4d\zeta d\overline{\zeta}(1+\zeta\overline{\zeta})^{-2}$ and $4d\eta d\overline{\eta}(1+\eta\overline{\eta})^{-2}$ while $\mathbb{Q}$ admits 
the holomorphic 2-form
\[\frac{d\zeta\wedge d\xi}{\eta}= 2\frac{d\zeta\wedge d\eta}{\zeta}=2\frac{d\eta\wedge d\xi}{\xi},\]
 so that we can calculate the Jacobians for the Gauss maps as in section 2.
 \begin{prop}
 
  For the future Gauss map we find
  \[
  \frac{4d\zeta d\overline{\zeta}}{(1+\zeta\overline{\zeta})^{2}}=\frac{1}{(t^a\ell_a)^2}(\rho^2-\sigma\overline{\sigma})dA,\]
  for the past Gauss map
  \[
  \frac{4d\eta d\overline{\eta}}{(1+\eta\overline{\eta})^{2}}=\frac{1}{(t^an_a)^2}(\rho'^2-\sigma'\overline{\sigma}')dA,\]
  and for the complex Gauss map
  \[
  \frac{d\zeta\wedge d\xi}{\eta}=(-\rho\rho'+\sigma\sigma')dA.\]
    
 \end{prop}
\noindent As in section 2, we integrate these expressions over $S$:
\begin{prop}{\bf Generalised Gauss-Bonnet Theorem}

 For the three cases treated above integration gives:
\be\label{41} \int_S\frac{1}{(t^a\ell_a)^2}(\rho^2-\sigma\overline{\sigma})dA=\int_S\frac{1}{(t^an_a)^2}(\rho'^2-\sigma'\overline{\sigma}')dA=\frac{4\pi}{t^at_a}\ee
\[ \int_S(-\rho\rho'+\sigma\sigma')dA=2\pi.\]
\end{prop}
\noindent In (\ref{41}) I have included the term $t^at_a$ explicitly both to give a slightly more general formula (valid when $t^a$ is any constant time-like vector) and to point up the resemblance 
to Newman's expression for the ${\mathcal{H}}$-space metric \cite{new1}.

\medskip

Next we turn to consideration of possible generalisations of the notion of parallel bodies and Propositions 2.2-2.4. For this we need to write down and solve the \emph{Sachs equations} which are the NP spin-coefficient 
equations for the evolution of $\rho,\sigma$ and $\rho', \sigma'$ \cite{np1} Given $S$ we consider the null hypersurface $\mathcal{N}$ generated by the outgoing null normals to $S$. We scale $\ell^a$ to be affinely 
parametrised:
\[D\ell^a\equiv \ell^b\nabla_b\ell^a=0,\]
and choose an affine parameter $s$ with
\[Ds=1,\;\;\;s=0\mbox{  at  }S.\]
Then the Sachs equations are
\[D\rho=\rho^2+\sigma\overline{\sigma},\;\;\;D\sigma=2\rho\sigma,\]
while the area element is carried along $\ell^a$ according to
\[D(dA)=-2\rho dA.\]
Then the Sachs equations can be solved explicitly as
\be\label{47}
\rho(s)=\Delta^{-1}(\rho_0-s(\rho_0^2-\sigma_0\overline{\sigma}_0)),\;\;\;\sigma(s)=\Delta^{-1}\sigma_0\ee
with
\[\Delta=1-2s\rho_0+s^2(\rho^2_0-\sigma_0\overline{\sigma}_0)\]
and $\rho_0,\sigma_0$ are the values at $S$. For the area element we similarly find
\[dA(s)=\Delta dA_0.\]
We deduce at once the following proposition:
\begin{prop}
 For a future convex surface, the outgoing null hypersurface encounters no caustics to the future.
\end{prop}
{\bf Proof: } Caustics to the future are signalled by singularities in $\rho(s)$, or equivalently zeroes in $\Delta$, for positive $s$ but from the definition of future convex $\Delta$ is positive definite in this range.

\medskip 

There is a corresponding statement for past convex.

\medskip

For the analogue of Proposition 2.2, Steiner's Theorem, we need to integrate $dA(s)$. However there is a problem of \emph{weights} in the GHP sense: at this point we have the freedom to rescale $\ell^a$ at $S$ by
\be\label{49}\ell^a\rightarrow \Omega(\theta,\phi)\ell^a,\ee
and under this transformation
\[s\rightarrow \Omega^{-1}s,\;\;\;\rho_0\rightarrow \Omega \rho_0,\;\;\;\sigma_0\rightarrow \Omega \sigma_0,\]
so that we would not get a formula like (\ref{11}) by simply integrating $dA(s)$. The simplest way to resolve this difficulty is to choose a constant unit time-like vector $t^a$ and define
\[\hat{s}=st^a\ell_a,\]
as $\hat{s}$ is unchanged by (\ref{49}) and now we can integrate $dA(s)$:
\begin{prop}Generalised Steiner Theorem
\[ A(\hat{s})=A_0+2\hat{s}\hat{M}+4\pi\hat{s}^2,\]
where
\be\label{53}\hat{M}=-\int_S\frac{\rho_0}{t^a\ell_a}dA\ee
and we have used Proposition 4.3.
\end{prop}
\noindent Now we might hope to prove a counterpart of the Brunn-Minkowski Theorem, Proposition 2.3, or of Minkowski's Inequality, Proposition 2.4. However this is impossible since with $\hat{M}$ 
as in (\ref{53}) examples can be found to show that it is not the case that $\hat{M}^2\geq 4\pi A$. In fact the Isoperimetric Inequality (\ref{16}) can also be violated in Minkowski space in the following sense: 
given a space-like 2-surface $S$ one may be 
able to find space-like 3-surfaces spanning $S$ on which the volume $V$ enclosed by $S$ and the area $A$ of $S$ have
\[36\pi V^2>A^3.\] 
The correct inequality to generalise (\ref{15}) would seem to be one proposed by Gibbons and Penrose in an investigation of Cosmic Censorship, \cite{rp1,g1}. This may be phrased as follows: consider 
the vector
\be\label{55}
P^a=\frac12\int_S(\rho'\ell^a-\rho n^a)dA=\int_Sp^adA,\ee
with $p^a=\frac12(\rho'\ell^a-\rho n^a)$. If $S$ is future 
convex with $\rho'>0$ as well, or past convex with $\rho<0$ as well, then one conjectures the inequality:
\be\label{56}
P_aP^a\geq 2\pi A,\ee
where $A$ is the area of $S$. (This is not the form in which the inequality is stated by Gibbons and Penrose but I believe it to be equivalent.) 

Note that the \emph{mean curvature vector} $H_a$ of $S$, equivalently the trace of the second fundamental form, is
\be\label{57}
H_a=\frac12(\rho'\ell_a+\rho n_a).\ee
The significance of $H_a$ is that, given a vector field $X^a$ on $S$, the rate of change of the area of $S$ under displacement along $X^a$ is
\[\dot{A}=\int_sH_aX^adA.\]
We see that the vector $p^a$ in (\ref{55}) lies in the normal 2-plane to $S$ and is orthogonal to $H_a$ so it defines the direction in which $dA$ does not change (to first order). Also, by taking $X^a$ to be a constant 
translation, under which the area will not change, it is clear that
\[\int_SH_adA=0\mbox{  and so  }\int_S\rho'\ell^adA=-\int_S\rho n^adA.\]
Thus we can write $P^a$ as
\[P^a=\int_S(-\rho)n^adA=\int_S\rho'\ell^adA.\]
In this form it is clear that $P^a$ is time-like and future pointing for past or future convex $S$.

\medskip

As partial confirmation of (\ref{56}) we note that if $S$ lies in a flat space-like 3-surface with unit time-like normal $t^a$ then
\[P^a=\frac{1}{\sqrt{2}}Mt^a,\]
with $M$ as in (\ref{13}). Thus in this case (\ref{56}) is Minkowski's inequality. Further, if $S$ lies in an in- or out-going null cone then (\ref{56}) can be established directly, 
as it reduces to an inequality for functions on the unit sphere, \cite{rp1,g1}, which can be proved \cite{t1,t2}. Finally one can verify (\ref{56}) for surfaces infinitesimally close to a round sphere in a flat hyperplane. 
What is still lacking is a proof of (\ref{56}) in full generality, subject only to the conditions of convexity\footnote{The claimed proof in \cite{g2} is defective: see e.g. \cite{mm}.}.


\end{document}